\documentclass[preprint,aps,showpacs,pre]{revtex4}
\usepackage{latexsym}
\usepackage{graphics}
\usepackage{amsmath}
\usepackage{graphicx}
\usepackage{amsfonts,amssymb}
\usepackage{amsthm}
\usepackage{epsfig}
\begin{document}

\title{The Transformation-Groupoid Structure of the q-Gaussian Family}
\author{Angel Akio Tateishi$^{1,2}$, Rudolf Hanel$^1$and  Stefan Thurner$^{1,3,4}$ 
}                     
\affiliation{$^1$ Section for Science of Complex Systems, Medical University of Vienna, Spitalgasse 23, 1090 Vienna, Austria  \\
$^2$ Departamento de F\'isica, Universidade Estadual de Maring\'a, Avenida Colombo, 5790 -- 87020-900 Maring\'a - PR, Brazil \\
$^3$ Santa Fe Institute, 1399 Hyde Park Road, Santa Fe, NM 87501, USA \\
$^4$ IIASA, Schlossplatz 1, A-2361 Laxenburg, Austria}
%
%
%

\begin{abstract}
The $q$-Gaussian function emerges naturally in various applications of statistical mechanics of non-ergodic and complex systems. 
In particular it was shown that in the theory of binary processes with correlations, the $q$-Gaussian can appear as a 
limiting distribution. Further, there exist several problems and situations where, depending on procedural or algorithmic details of data-processing, 
$q$-Gaussian distributions may yield distinct values of $q$, where one value is larger, the other smaller than one. 
To relate such pairs of $q$-Gaussians it would 
be convenient to map such distributions onto one another, 
ideally in a way, that any value of $q$ can be mapped uniquely to any other value $q'$. 
So far a (duality) map from 
$q\to q'=\frac{7-5q}{5-3q}$ 
was found, mapping $q$ from the interval $q\in [-\infty, 1] \to q'\in [1, 5/3]$. Here we complete   
the theory of transformations of $q$-Gaussians by deriving a general map $\gamma_{qq'}$, that transforms normalizable 
$q$-Gaussian distributions onto one another for which $q$ and $q'$ are in the range of $[1,\,3)$. 
By combining this with the previous result, a mapping from any value of 
$q \in [-\infty,3)$ is possible to any other value $q'\in [-\infty,3)$. 
We  show that the action of $\gamma_{qq'}$ on the set of $q$-Gaussian distributions is a 
transformation groupoid. 
\end{abstract}
\maketitle

\section{Introduction}

The $q$-Gaussian is a fundamental power-law probability distribution function defined as
\begin{equation}\label{qGaussian}
G_{q}(x)\equiv \frac{1}{Z_{q}}[1-(1-q)\beta x^{2}]^{\frac{1}{1-q}},
\end{equation}
where $Z_{q}$ is the normalization, and $\beta>0$ is some constant.
In the limit $q\to 1$ $q$-Gaussians converge to the Gaussian distribution. 
In the range $q\in[-\infty,1)$ the function $G_q(x)$ is defined only for $x\in[-L,L]$, with $L=1/\sqrt{1-q}$, i.e.
for $|x|>L$ one has to define $G_q(x)\equiv 0$. For $q>1$, $G_q(x)$ is defined for all $x\in[-\infty,\infty]$. 
Note, that for $q$ in the range $[-\infty, 5/3]$, the existence of a second moment, while 
for $q$ in $[-\infty, 3)$, normalization is ensured for the $q$-Gaussian distribution.

$q$-Gaussians can be derived in various ways and play a fundamental role in generalized statistical mechanics. 
$q$-Gaussians (and $q$-exponentials) naturally appear as a particular class of statistics in systems, 
where one of the  four Shannon-Khinchin (SK) \cite{Shannon,Khinchin} axioms\footnote{
	Shannon-Khinchin axioms:
	(i) Entropy is a continuous function of the probabilities $p_i$ only, i.e. $s$ should not explicitly depend on 
	any other parameters. 
	(ii) Entropy is maximal for the equi-distribution $p_i=1/W$.
	-- From this the concavity of $s$ follows. 
	(iii) Adding a state $W+1$ to a system with $p_{W+1}=0$ does not change the entropy of the system. 
	-- From this $s(0)=0$ follows.
	(iv) Entropy of a system composed of 2 sub-systems $A$ and $B$, is $S(A+B)=S(A)+S(B |A)$.
	\label{foo_shannon}
} 
-- the separation axiom -- 
is violated, e.g. in 
non-ergodic or non-Markovian systems \cite{HTclassification,HTextensive}. 
This class of statistics is sometimes referred to as $q$-statistics \cite{TsallisBook}, which has numerous 
applications in physical, biological and social systems.
Several investigations indicate that the frequent occurrence of $q$-Gaussian distributions in nature is not a mere coincidence. 
There may for instance exist deep connections with {\em large deviation theory} \cite{Ruiz2012,Touchette2009} 
for random variables of strongly correlated random processes. 
Also in the context of signal normalization, which is frequently required for transforming a signal to match 
the finite input range of an actual detector, it has been shown that some normalization procedures affect signal properties \cite{vignat2009}. 
In particular, for some classes of stochastic signals there exist two simple normalization procedures such that 
the distribution of the normalized signal received by the detector is is $q$-Gaussian. For one of the two procedures the resulting $q$-values are always 
$q>1$ while for the other $q<1$ \cite{vignat2009}.

Another situation where $q$-Gaussians naturally appear is in the context of 
binary correlated processes \cite{rio}. 
A large class of correlated binary processes is completely determined by its limit distribution, i.e. by the distribution of the number of counts of 
heads $n_+$ and tails $n_-$, as the number of tosses $N$ goes to infinity. 
To study limit distributions of such processes it is necessary to fix a normalization scheme for these numbers of counts. 
Here again two ``natural'' scaling procedures exist. In one case one normalizes the number of counts for heads and tails simply by 
dividing by the number of tosses. The difference $x=(n_+-n_-)/N$ of these normalized counts therefore is a random variable $x \in [-1,1]$. 
Alternatively, one can ``zoom in'' on the distribution function of $x$ and study another random variable $y \propto x/\sqrt{1-x^2}$ with $y\in[-\infty,\infty]$. 
If a correlated binary process has a $q$-Gaussian as a limit distribution of $x$ with $-\infty < q\leq 1$, 
then the limit distribution of $y$ is another $q$-Gaussian with $5/3>q'=f(q)\geq 1$. 
The function $f(q)=\frac{7-5q}{5-3q}$ turns out to be a duality, i.e. $f(f(q))=q$ on $[-\infty, 5/3]$.
Since the two $q$-Gaussians are generated by the same correlated binary process they map onto one another in a natural way.

In conclusion, there arise situations where two $q$-Gaussians with distinct values of $q$ are generated by a common 
underlying process. Yet, a general transformation to map any two $q$-Gaussians with values $q$ and $q'$ in $[-\infty, 3)$ onto one another 
was hitherto still missing.
The $q$-Gaussian might also be naturally related to a generalization of the Fourier-transform \cite{gell}.

Despite its numerous applications, the mathematical properties of the $q$-Gaussian distribution functions are not yet fully explored. 
Recently symmetries were found and described in generalized statistical mechanics \cite{HTG,HTG2}. 
In this spirit here we will explore the group -- or more precisely -- the {\it groupoid} structure of the $q$-Gaussian distribution family. 
We will show that there exists a family of transformations that map one $q$-Gaussian with a specific value of $q\in[1,3)$ onto 
another $q$-Gaussian (with $q'\in[1,3)$ and $q'\neq q$) and prove that this family  satisfies the properties of a groupoid.
As a particular application of these findings, in combination with the results from \cite{rio}, then allows us
to extend the transformation groupoid to maps $q\to q'$ with both $q$ and $q'$ in $[-\infty,3)$.
As a result we can represent any possible relation between all normalizable $q$-Gaussians.

This paper is organized as follows. In Section \ref{sec:1}, we establish a set $\mathcal{G}$ of scaling functions $\gamma_{q'q}$ that 
map $q$-Gaussians onto $q$-Gaussians for $q\in[1,3)$. In Section \ref{sec:2}, we prove that the action of $\gamma_{q'q}$ on $q$-Gaussians forms a transformation groupoid, 
$(\mathcal{G},\circ)$, under a specific  composition operation $\circ$. Finally, in Section \ref{sec:3}, we conclude.


\section{The set of scaling functions}
\label{sec:1}
A way to map one normalizable $q$-Gaussian distribution onto another is to 
identify their probabilities, i.e. if $y$ and $y'$ are $q$-Gaussian distributed random variables, with $q$-values $q$ and $q'$ respectively, 
then we ask which values of  $z$ and $z'$ allow to equate the probabilities $P(z>y>0)=P(z'>y'>0)$. 
We achieve this by determining the scaling functions $\gamma(z)$ such that  
\begin{equation}\label{map1}
\int^{\gamma(z)}_{0}dy~ G_{q}(y) = \int^{z}_{0}dy'~G_{q'}(y'),
\end{equation}
where $G_{q}(y)$ and $G_{q'}(y')$ are two $q$-Gaussian distributions, defined in Eq. (\ref{qGaussian}).
To solve this equation it is suitable to consider its differential form,  
\begin{equation}\label{map2}
\frac{d \gamma(z)}{dz} G_{q}(\gamma(z))=G_{q'}(z),
\end{equation}
with  $\gamma(0)=0$, and $z\in [-\infty,+\infty]$. The solutions of this differential equation can be identified with a map 
$\gamma_{q'q}: G_{q} \rightarrow G_{q'}$ such that  $G_{q'}(z)\equiv \gamma_{q'q}'(z)G_{q}(\gamma_{q'q}(z))$. 
By using the definition of $q$-Gaussians (with $\beta=1$ for simplicity) and  separation of variables, we write Eq. (\ref{map2}) as 
\begin{equation}\label{map3}
\frac{1}{Z_{q}}\int d\gamma(z) [1-(1-q)\gamma^{2}(z)]^{\frac{1}{1-q}} = \frac{1}{Z_{q'}}\int dz [1-(1-q')z^{2}]^{\frac{1}{1-q'}}.
\end{equation}
Note that both integrals in Eq. (\ref{map3}) have the same structure and can be expressed in terms of hypergeometric functions 
\begin{equation}\label{map4}
\frac{1}{Z_{q}}  \gamma(z)~  _{2}F_{1}\left( \frac{1}{2}, \frac{1}{q-1}; \frac{3}{2}; -(q-1)\gamma^{2}(z)   \right)     = 
  \frac{1}{Z_{q'}}z~  _{2}F_{1}\left( \frac{1}{2}, \frac{1}{q'-1}; \frac{3}{2}; -(q'-1)z^{2}   \right).
\end{equation}
 To determine $\gamma(z)$ we can use the relations \cite{TableGrad}
\begin{equation}
_{2}F_{1}\left(\alpha,\beta;\gamma;z\right) = (1-z)^{-\alpha}~  _{2}F_{1}\left(\alpha,\gamma-\beta;\gamma;\frac{z}{z-1}\right)
\end{equation}
and
\begin{equation}
B_{x}(m,n) = \int_{0}^{x} t^{m-1}(1-t)^{n-1}dt = \frac{x^{m}}{m}~ _{2}F_{1}(m,1-n;m+1;x).
\end{equation}
It is then easy to show that Eq. (\ref{map4}) may be written as
\begin{equation}
\label{eq8}
 \frac{1}{Z_{q}\sqrt{q-1}} B_{\frac{(q-1)\gamma^{2}(z)}{1+(q-1)\gamma^{2}(z)}} \left( \frac{1}{2}; \frac{1}{q-1}-\frac{1}{2} \right) =  \frac{1}{Z_{q'}\sqrt{q'-1}}
  B_{\frac{(q'-1)z^{2}}{1+(q'-1)z^{2}}} \left( \frac{1}{2}; \frac{1}{q'-1}-\frac{1}{2} \right).
\end{equation}
\begin{figure}[tb]
\begin{center}
	\includegraphics[scale=0.42]{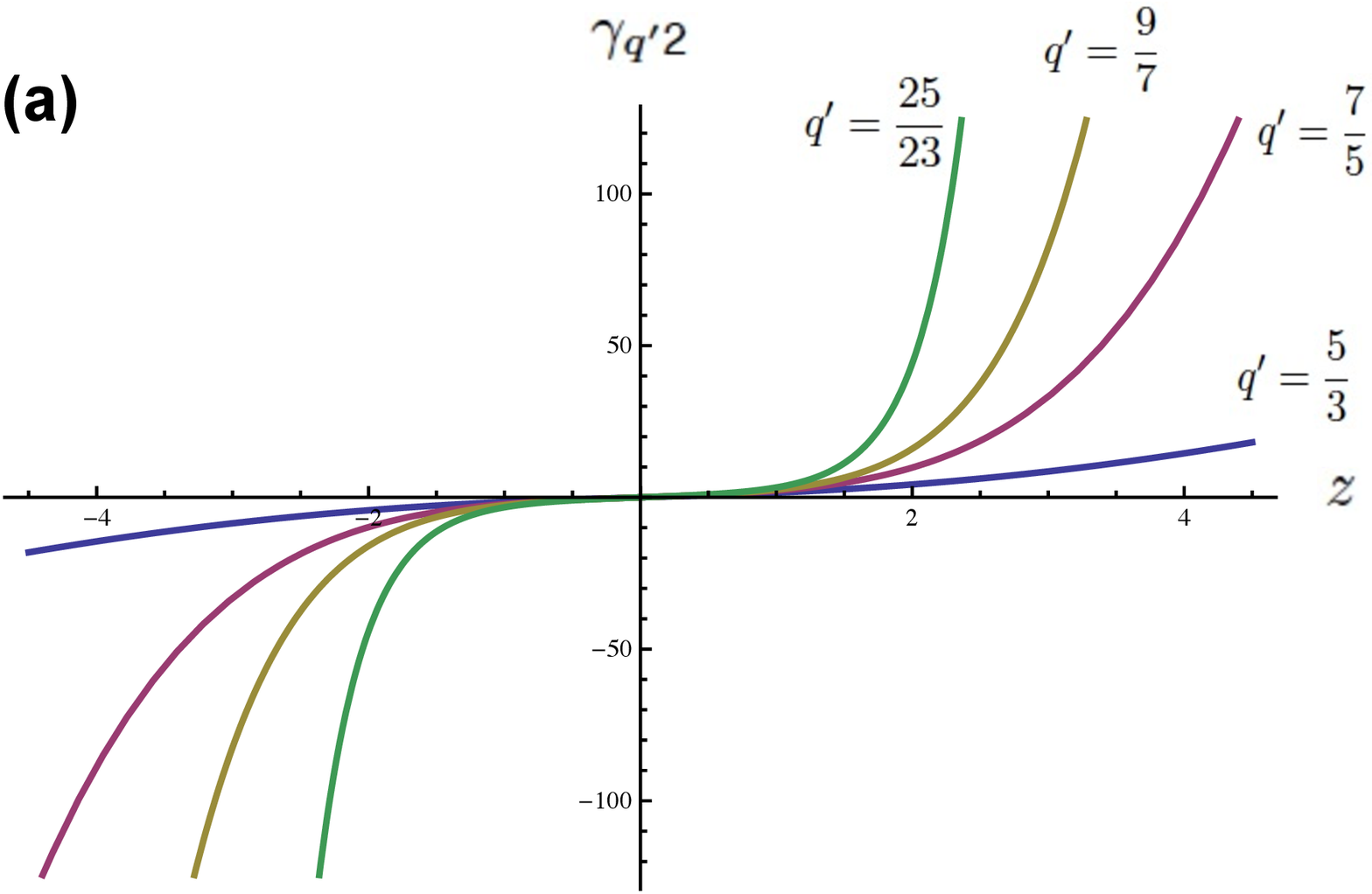}\quad 
	\includegraphics[scale=0.42]{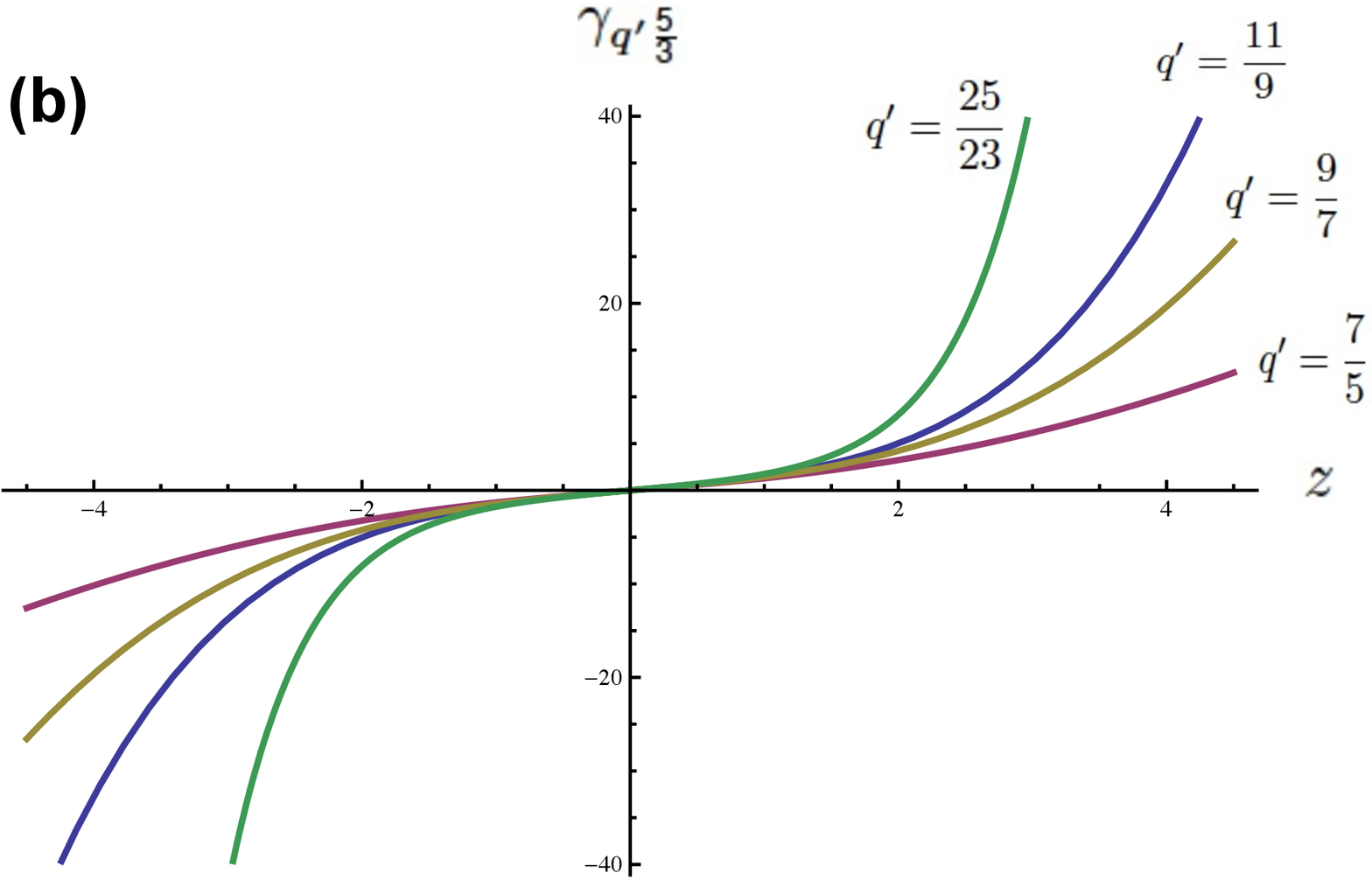}
\caption{(a) 
Scaling functions $\gamma_{q'2} : G_{2}\rightarrow G_{q'}$ for values $q' = \lbrace \frac{5}{3},\frac{7}{5},\frac{9}{7},\frac{25}{23} \rbrace $. 
(b) Scaling functions 
$\gamma_{q'\frac{5}{3}} : G_{\frac{5}{3}}\rightarrow G_{q'}$ for values $q' = \lbrace \frac{7}{5},\frac{9}{7},\frac{11}{9},\frac{25}{23} \rbrace $.}
 \label{Fig:1}
\end{center}
\end{figure}

We assume from now on that $1\leq q<3$. 
The normalization $Z_{q}=\int_{-\infty}^{\infty}dx ~G_{q}(x)$ is given by
\begin{equation}\label{normFactor}
Z_{q} = \frac{1}{\sqrt{q-1}}B\left( \frac{1}{2}, \frac{1}{q-1}-\frac{1}{2} \right).
\end{equation}
Using the definition of the regularized incomplete Beta-function, $\mbox{\large{I}}_{x}(a,b) \equiv B_{x}(a,b)/B(a,b), $ with $a>0$, $b>0$, 
and $x\in[0,1]$, we write Eq. (\ref{eq8}) as 
\begin{equation}
\mbox{\Large{I}}_{\frac{(q-1)\gamma^{2}(z)}{1+(q-1)\gamma^{2}(z)}}\left( \frac{1}{2}~; \frac{1}{q-1}-\frac{1}{2} \right) = 
\mbox{\Large{I}}_{\frac{(q'-1)z^{2}}{1+(q'-1)z^{2}}}\left( \frac{1}{2}~; \frac{1}{q'-1}-\frac{1}{2} \right)\,.
\end{equation}
Since $y=I_x(a,b)$ is monotonically increasing in $x$, there exists the inverse function $x=I^{-1}_y(a,b)$. Using this
we can finally obtain the map 
\begin{equation}\label{The Thing}
\gamma_{q'q}(z) = \pm \frac{1}{\sqrt{q-1}}  \left(\frac{\mbox{\Large{I}}^{-1}_{\mbox{\large{I}}_{\frac{(q'-1)z^{2}}{1+(q'-1)z^{2}}}\left( \frac{1}{2}~; \frac{1}{q'-1}-\frac{1}{2} \right)  }
 \left( \frac{1}{2}~; \frac{1}{q-1}-\frac{1}{2} \right) }{1- \mbox{\Large{I}}^{-1}_{\mbox{\large{I}}_{\frac{(q'-1)z^{2}}{1+(q'-1)z^{2}}}\left( \frac{1}{2}~; \frac{1}{q'-1}-\frac{1}{2} \right)  }
  \left( \frac{1}{2}~; \frac{1}{q-1}-\frac{1}{2} \right) }  \right)^{\frac{1}{2}} .
\end{equation}

\begin{figure}
 \centering
\includegraphics[scale=0.55]{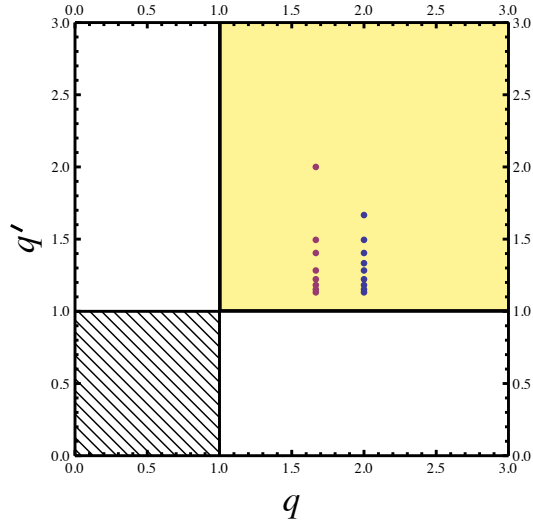}
\caption{The range of $1<q<3$ and $1<q'<3$  define the region in which $G_{q}$ and $G_{q'}$ can be mapped to each other by $\gamma_{q'q}\in\mathcal{G}$. 
The points  represent the particular cases listed in Table I.}
\label{Fig:2}
\end{figure}

These maps  $\gamma_{q'q} : G_{q}\rightarrow G_{q'}$  form the family $\mathcal{G}$ of scaling transformations that act on 
the set of $q$-Gaussian distributions through Eq. (\ref{map1}). Remarkably, for some particular values of $q$ and $q'$, Eq. (\ref{The Thing}) 
can be simplified and expressed in terms of simple functions.  In Table \ref{tab:1}  we show some examples of these particular cases. 
In Fig. \ref{Fig:1}  some of these maps are plotted.  In Fig. \ref{Fig:2} we show the region for $q$ and $q'$ that can be mapped to each other through $\gamma_{q'q}$. 
The particular cases listed in Table I are plotted as points.

\begin{table}
\caption{Particular cases of $\gamma_{qq'}$ for which simple functions can be found.}
\label{tab:1}     
\centering

\begin{tabular}{lll}

\hline\noalign{\smallskip}
Transformation:  $G_{q}(z) \rightarrow G_{q'} (z)$& ~   Scaling function: $\gamma_{q'q}(z)$   \\
\noalign{\smallskip}\hline\noalign{\smallskip}
$G_{2}(z) \rightarrow G_{5/3} (z)$& ~       $ \tan\left( \frac{\pi z}{(6+4z^{2})^{1/2}} \right)$  \\
$G_{2}(z) \rightarrow G_{7/5}(z) $& ~       $ \tan\left( \frac{\pi z (15+4z^{2})}{2\sqrt{2} (5+2z^{2})^{3/2} }   \right)$\\
$G_{2}(z) \rightarrow G_{9/5}(z) $& ~        $ \tan\left( \frac{\pi z ( 735 +280z^{2} + 32z^{4}) }{8\sqrt{2} (7+2z^{2})^{5/2}}  \right)$    \\
$G_{2}(z) \rightarrow G_{11/9}(z) $& ~        $ \tan\left( \frac{\pi z (25515 + 4z^{2}(2835 + 504z^{2} + 32z^{4} )  )  }{  16\sqrt{2} (9+2z^{2})^{7/2}  }  \right)$ \\
$G_{2}(z) \rightarrow G_{13/9}(z) $ & ~        $ \tan\left( \frac{\pi z (4 611 915 +16 z^{2}(139755+4z^{2}(7623+792z^{2}+32z^{4}))) }{128\sqrt{2}(11+2z^{2})^{9/2}}  \right)$    \\ 
$G_{2}(z) \rightarrow G_{q'\rightarrow 1}(z) $ & ~ \ \ \ \ \ \ \ \ \ \ \ \ \ \ \ \ \ \ \ \ \   \vdots~\\ 
 $G_{5/3}(z) \rightarrow G_{2} (z)$& ~       $\frac{\sqrt{6} \arcsin\left(\frac{z}{\sqrt{1+z^{2}}}\right)}{\pi^{2}- 4\arcsin^{2}\left(\frac{z}{\sqrt{1+z^{2}}}\right)}$     \\
 $G_{5/3}(z) \rightarrow G_{7/5}(z) $& ~     $ \frac{\sqrt{3}z(11+4z^{2})}{\sqrt{500+150z^{2}}}$      \\ 
 $G_{5/3}(z) \rightarrow G_{9/5}(z) $& ~        $\frac{\sqrt{3}z(735+280z^{2}+32z^{4})}{\sqrt{1075648 +456190z^{2}  +54880z^{4}}}$   \\
  $G_{5/3}(z) \rightarrow G_{11/9}(z) $& ~       $\frac{z(25515 + 4z^{2}(2835+504z^{2}+32z^{4}))}{27\sqrt{6}\sqrt{93312+45927z^{2}+8568z^{4}+560z^{6}}}$ \\ 
  $G_{5/3}(z) \rightarrow G_{13/9}(z) $ & ~          $ \frac{\sqrt{\frac{3}{22}}z (4611915+16z^{2}(139755+4z^{2}(7623+792z^{2}+32z^{4})))}{121\sqrt{ 119939072+3z^{2}
 (21398487+32z^{2}(152 823+16324z^{2}+672z^{4})) }}$ \\
$G_{5/3}(z) \rightarrow G_{q'\rightarrow 1}(z) $ & ~ \ \ \ \ \ \ \ \ \ \ \ \ \ \ \ \ \ \ \ \ \   \vdots~\\
\noalign{\smallskip}\hline
\end{tabular}
\end{table}

\section{Transformation Groupoid}
\label{sec:2}
In this section we prove that the family $\mathcal{G}=\{\gamma_{q'q}\}_{q,q'\in[1,3)}$,
obtained in the previous section, has special 
properties that allow us to classify $(\mathcal{G},\circ)$ as a {\it groupoid} under the composition 
operation
\begin{equation}\label{compmap}
\gamma_{q''q'}\circ \gamma_{q'q}(z) \equiv \gamma_{q'q}(\gamma_{q''q'}(z)).
\end{equation}
As pointed out in \cite{groupoid1}, this  algebraic structure, in which every  element of the  set is invertible  (formally, every morphism is an 
isomorphism), is an extension  of the concept of groups, for more details see \cite{groupoid1,groupoid2,groupoid3}. 
Here, we focus on those conditions 
constituting a transformation groupoid.

Let us consider the action of the set of maps $\mathcal{G}$ on the set of $q$-Gaussian distributions $\mathcal{Q}=\{G_{q}\}_{q\in[1,3)}$ (elements). 
There exist two functions $s:\mathcal{G}\to \mathcal{Q}$ and $t:\mathcal{G}\to \mathcal{Q}$ that can be used to assign two unique elements 
to each map $\gamma\in \mathcal{G}$ with $\gamma=\gamma_{q'q}$ different from the identity map. 
One element $s(\gamma_{q'q})=G_{q}$ is the ``source'', the other
element $t(\gamma_{q'q})=G_{q'}$ is the ``target'' of the map $\gamma_{q'q}$ (i.e. 
the right $q$-index of $\gamma_{q'q}$ indicates the source, the left the target). 
This action may be expressed as $\gamma_{q'q} : G_{q}\rightarrow G_{q'}$. 
Therefore $(\mathcal{G},\circ)$ forms a transformation groupoid on the family of $q$-Gaussians with $q\in [1,3)$,
where the binary operation satisfies the properties of (i) composition, (ii) associativity, (iii) the existence of an unique identity element, 
and (iv) the existence of inverse elements.


\begin{enumerate}
  \item {\bf Composition}. The composition of two elements in a groupoid is defined if and only if 
  the target of the one map coincides with the source of the other map and is given by 
  Eq. (\ref{compmap}).

The idea of the proof is to show that if $\gamma_{q''q'}$ and $\gamma_{q'q}$ are solutions of Eq. (\ref{map1}), then their composition $\gamma_{q''q'}\circ\gamma_{q'q}$ is the solution $\gamma_{q''q}$ of Eq. (\ref{map1}).
\begin{proof}
Let $\gamma_{q''q'}(z)=f(z)$ and $\gamma_{q'q}(y)=g(y)$. From Eq. (\ref{map2}) we have that $f(z)$ and $g(y)$ are solutions of 
\begin{equation}\label{eqHT}
f'(z)G_{q'}(f(z)) = G_{q''}(z)
\end{equation}
 and
 \begin{equation}\label{eqHT2}
 g'(y)G_{q}(g(y)) = G_{q'}(y).
\end{equation}
Consider $f\circ g(y)=g(f(z))$. Inserting $y =  f(z)$ in Eq. (\ref{eqHT2}) yields
 \begin{equation}\label{eqHT3}
 g'(f(z))G_{q}(f(z)) = G_{q'}(f(z)).
 \end{equation}
Multiplying both sides of Eq. (\ref{eqHT3}) with $f'(z)$,  
  \begin{equation}
  \underbrace{g'(f(z))f'(z) }_{(g(f(z)))'}G_{q}(g(f(z))) = \underbrace{f'(z)G_{q'}(f(z))}_{\mbox{right-hand side of Eq. (\ref{eqHT})}}\,,
  \end{equation}
and using $f\circ g(z)=g(f(z))$ on the left side of the equation finally yields  
  \begin{equation}
  (f\circ g(z))'G_{q}(f\circ g(z))=G_{q''}(z)\,,
  \end{equation}
  and the composition $\gamma_{q''q'}\circ\gamma_{q'q}(z)$  is the map $\gamma_{q''q}: G_{q}\rightarrow G_{q''}$ .  \ 
       \end{proof} 
 In Fig. \ref{Fig:3}a, the composition is depicted graphically. This  result  will be useful to prove the next three properties.


 \begin{figure}[tb]
	\includegraphics[scale=0.4]{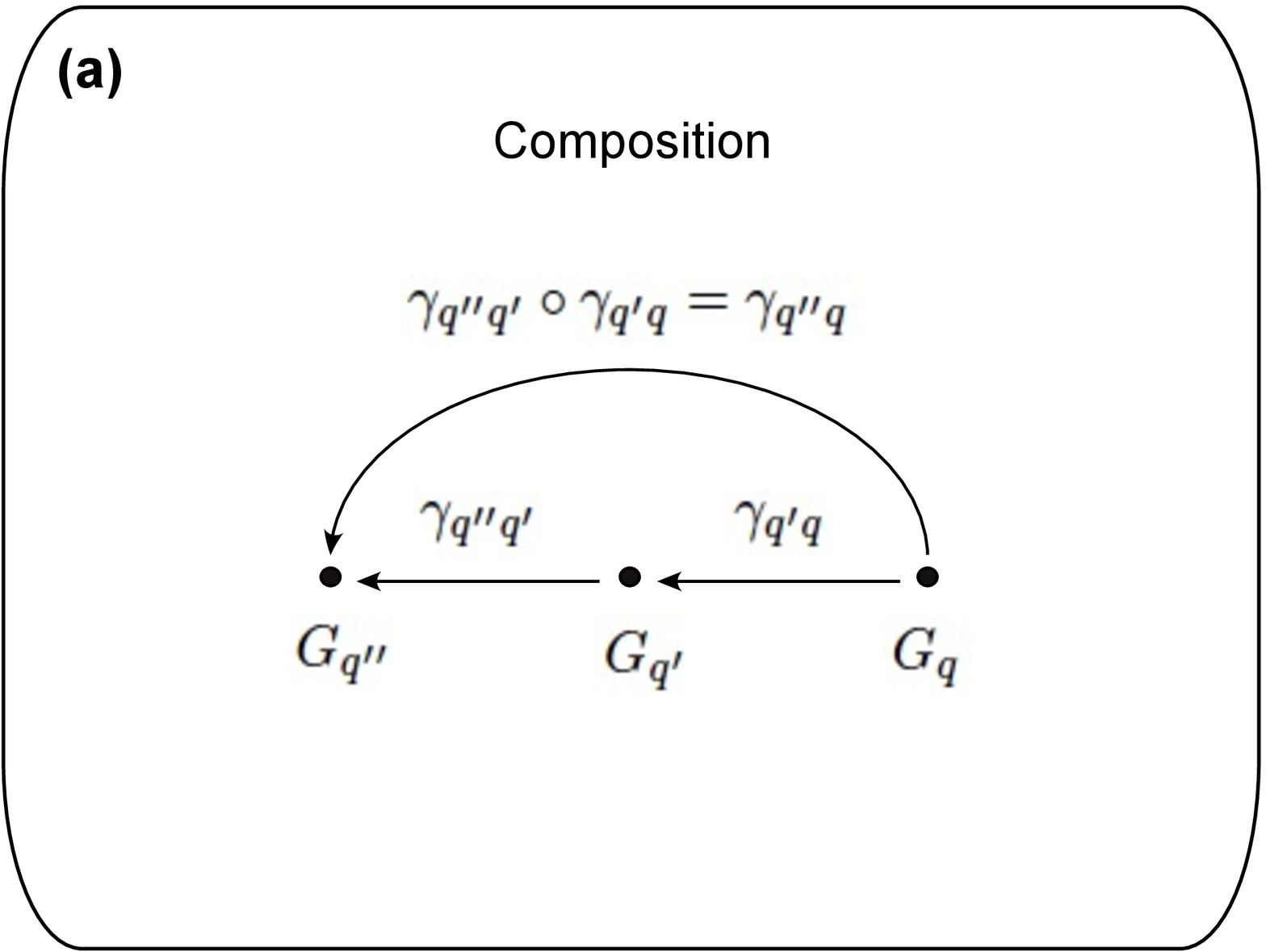} 
	\includegraphics[scale=0.4]{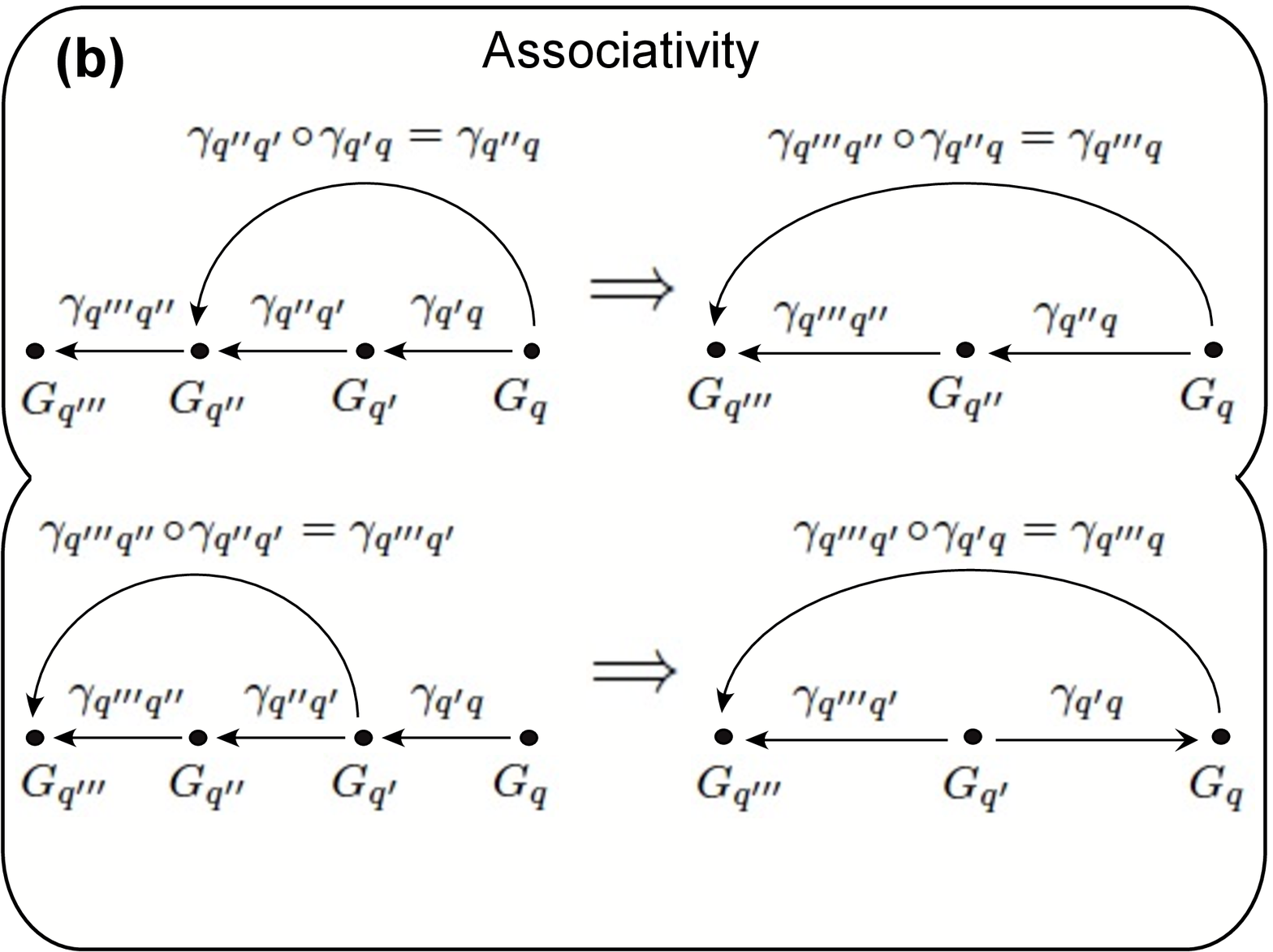} 
       \includegraphics[scale=0.4]{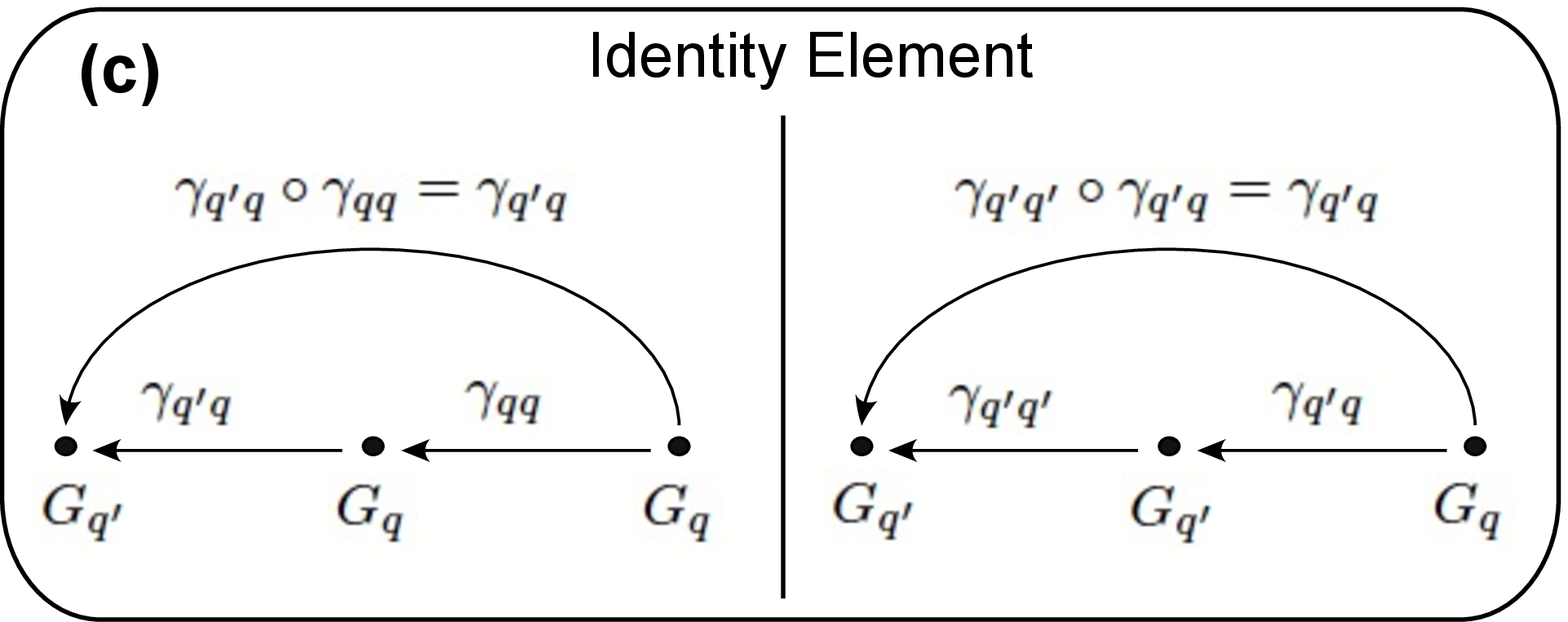}
         \includegraphics[scale=0.4]{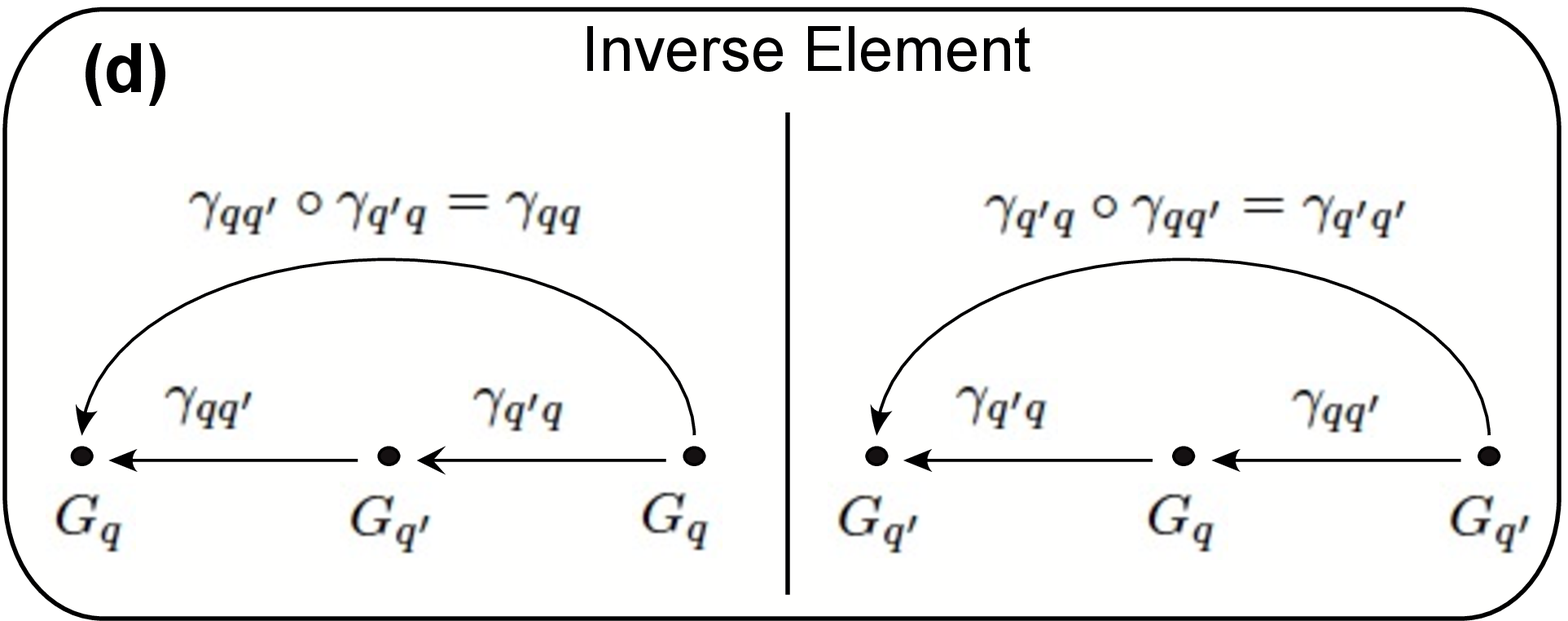}
\caption{(a) Composition operation (multiplication) of the groupoid. (b) The upper panel illustrates the operation 
$ \gamma_{qq'}\circ(\gamma_{q'q''}\circ\gamma_{q''q'''})$ 
while the lower panel shows the operation $ (\gamma_{qq'}\circ\gamma_{q'q''})\circ\gamma_{q''q'''}$. 
Associativity means that the order of composition does not affect 
the final result. (c) The composition of the identity element ($\gamma_{qq}(z)=z, \forall q$) with another element maps the latter onto itself. 
The left and right panels illustrate left and right multiplication with the identity element. 
(d) Composition of an element with its existing inverse element yield the identity element. }
 \label{Fig:3}
\end{figure}
 

 \item {\bf Associativity}. 
 The composition rule $\circ$ is associative, i.e.
 \begin{equation}\label{assoc}
 \gamma_{q'''q''}\circ(\gamma_{q''q'}\circ\gamma_{q'q}) = (\gamma_{q'''q''}\circ\gamma_{q''q'})\circ\gamma_{q'q}. 
 \end{equation}

\begin{proof}
This property is a natural consequence of the previous one.
Applying the composition rule first to the terms within the parentheses on both sides of Eq. (\ref{assoc}) and then once more to the results
of both sides,
\begin{equation}
\underbrace{\gamma_{q'''q'}\circ \gamma_{q'q}}_{\gamma_{q'''q}} = \underbrace{\gamma_{q'''q''}\circ\gamma_{q''q}}_{ \gamma_{q'''q}},
\end{equation}
shows that the left and right side of  Eq. (\ref{assoc}) are in fact identical.
\hfill\hspace{1cm} 
\end{proof}
 In Fig. \ref{Fig:3}b, associativity is demonstrated graphically.


\item {\bf Identity element}.  There exists an unique identity element, ${\rm id}\in\mathcal{G}$, with ${\rm id}(z)=z$. 
Moreover for any $q\in[1,3)$ we have that $\gamma_{qq}={\rm id}$.

\begin{proof}
First note that $f\circ {\rm id}(z)={\rm id}(f(z))=f(z)$, and ${\rm id}\circ f(z)=f({\rm id}(z))=f(z)$ and therefore
${\rm id}(z)=z$, is indeed the identity element with respect to $\circ$. Using the definition of the composition rule one gets
$\gamma_{q'q}\circ\gamma_{qq}(z)=\gamma_{q'q}(z)$
and
$\gamma_{q'q'}\circ\gamma_{q'q}(z)=\gamma_{q'q}(z)$.
Moreover, by the definition of $\gamma_{q'q}$, Eq. (\ref{map1}), it directly follows that $\gamma_{qq}(z)=z$ for all $q$ and therefore
$\gamma_{qq}\equiv{\rm id}$ for all $q\in[1,3)$. 
\hfill\hspace{1cm} 
\end{proof}
 In Fig. \ref{Fig:3}c, the source and target identity elements are depicted graphically.


\item {\bf Inverse element}. For each map $\gamma_{q'q}\neq {\rm id}$,  
there exists a unique inverse element  $\gamma^{-1}_{q'q}=\gamma_{qq'}$ for which 
$\gamma_{qq'}\circ\gamma_{q'q}={\rm id}$ and $\gamma_{q'q}\circ\gamma_{qq'}={\rm id}$, for any $q\neq q'$ in $[1,3)$.

\begin{proof}
Using the composition rule immediately shows that $\gamma_{qq'}\circ\gamma_{q'q}=\gamma_{qq}$, for any $q'$ and $q$ in $[1,3)$. 
Since we already know that $\gamma_{qq}={\rm id}$, this completes the proof. 
\hfill\hspace{1cm} 
\end{proof}
 In Fig. \ref{Fig:3}d, the inverse element is depicted graphically.

\end{enumerate}


\section{Conclusions}
\label{sec:3}

In this work we have derived the transformation groupoid acting on $q$-Gaussian distribution functions with $q\in[1,3)$ by studying 
probability-preserving scaling transformations between $q$-Gaussians. We explicitly have shown that a groupoid structure allows to 
map any two $q$-Gaussian distribution functions with $1\leq q<3$ onto each other. 
It was shown in \cite{rio} that $q$-Gaussian distribution functions with compact support, $q\in[-\infty,1]$, can be mapped to $q$-Gaussians 
with finite second moments, i.e. $q'\in[1,5/3]$.  The particular map was shown to be $q\to q'=\frac{7-5q}{5-3q}$ which preserves probabilities 
in the same sense as the scaling transformations of the groupoid in this paper. 
Using this result \cite{rio} in combination with the main result of this present paper  it becomes immediately clear that the groupoid structure 
naturally extends to all normalizable $q$-Gaussian distribution functions ($q\in[-\infty,3)$). 
This is trivially seen by composing the groupoid elements derived in this work with the particular map derived in \cite{rio}. 
We therefore present a complete theory of mappings between normalizable $q$-Gaussians, 
allowing to relate any two normalizable $q$-Gaussian distribution functions and therefore to specify any relation between 
normalizable $q$-Gaussians that can be encountered.

\section*{ACKNOWLEDGMENTS}

We thank C. Tsallis for fruitful discussions. A.A.T. is grateful to CAPES for financial support under Grant No. 1507-12-5.

\end{document}